\documentclass[aps,preprint,superscriptaddress]{revtex4-1}

\usepackage{amssymb}
\usepackage{latexsym}
\usepackage{amsfonts}
\usepackage{graphicx}
\usepackage{epsfig}
\usepackage{amsmath}
\usepackage{float}
\usepackage{verbatim}
\usepackage{subfigure}
\usepackage{color}

\begin{document}

\title{Percolation of Partially Interdependent Scale-Free Networks}

\author{Di Zhou}
\affiliation{Center for Polymer Studies and Department of Physics,
Boston University, Boston, MA 02215 USA}

\author{Jianxi Gao}
\affiliation{Department of Automation, Shanghai Jiao Tong
University, 800 Dongchuan Road, Shanghai 200240, PR China}
\affiliation{Center for Polymer Studies and Department of Physics,
Boston University, Boston, MA 02215 USA}

\author{H. Eugene Stanley}
\affiliation{Center for Polymer Studies and Department of Physics,
Boston University, Boston, MA 02215 USA}

\author{Shlomo Havlin}
\affiliation{Department of Physics, Bar-Ilan University, Ramat-Gan 52900, Israel}

\date{02-18-2012}

\begin{abstract}

We study the percolation behavior of two interdependent scale-free
(SF) networks under random failure of 1-$p$ fraction of nodes. Our
results are based on numerical solutions of analytical expressions and 
simulations. We find that as the coupling strength between
the two networks $q$ reduces from $1$ (fully coupled) to $0$ (no
coupling), there exist two critical coupling strengths $q_1$ and
$q_2$, which separate three different regions
with different behavior of the giant component as a function of $p$.
(i) For $q \geq q_1$, an abrupt
collapse transition occurs at $p=p_c$. (ii) For
$q_2<q<q_1$, the giant component has a hybrid transition combined of both, abrupt decrease at
a certain $p=p^{jump}_c$
followed by a smooth decrease to zero for
$p < p^{jump}_c$ as $p$ decreases to zero.
(iii) For $q \leq q_2$, the giant component has a continuous second-order transition (at
$p=p_c$). We find that $(a)$ for $\lambda \leq 3$, $q_1 \equiv 1$;
and for $\lambda > 3$, $q_1$ decreases with increasing $\lambda$.
$(b)$ In the hybrid transition, at the $q_2 < q <
q_1$ region, the mutual giant component $P_{\infty}$ jumps discontinuously at $p=p^{jump}_c$
to a very small but non-zero value, and when reducing $p$,
$P_{\infty}$ continuously approaches to $0$ at $p_c = 0$ for $\lambda < 3$ and at $p_c > 0$ for $\lambda > 3$.
Thus, the known theoretical $p_c=0$
for a single network with $\lambda \leqslant 3$ is expected to be valid also for strictly partial
interdependent networks.

\end{abstract}


\maketitle

\newpage

\section{Introduction}
Complex networks appear in almost every aspect of science and
technology \cite{Watts1998,SF1999, Havlin2010,NewmanBook,
Caldarelli2007,Albert2002,Cohen2000,Callaway2001, DogBook,Song2005,Bashan2012,Barrat2008,Peixoto2012, Zeng2012}. 
An important
property of a network is its robustness in terms of node and link
failures. The robustness of a network is usually characterized by
the value of the critical threshold analyzed by percolation theory.
Recently, motivated by the fact that modern infrastructures are
significantly coupled together, the robustness of interdependent
networks has been studied \cite{Rinaldi2001, Laprie2007, Panzieri2008, Rosato2008,
Vespignani2010, Goldenberg2005, Di2012, Morris2012, Zhao2012, Leicht2009, Wang2013, Schneider2013, Pocock2012, Cho2010}. 
In interdependent networks, the failure
of nodes in one network generally leads to failure of dependent
nodes in other networks, which in turn may cause further damage to
the first network, leading to cascading failures and catastrophic
consequences.

The structure of complex networks is frequently non-homogeneous with a broad degree distribution.
In many cases, the degree distribution obeys a power-law form,
and the networks are called scale-free (SF) \cite{SF1999}. Real
networks that have been found to be well approximated by power-law degree
distribution, include between many others, the Internet, airline networks, protein
regulatory networks, and research collaboration networks \cite{SF1999, Albert2002, Caldarelli2007}.
Thus, the analysis of
interdependent scale-free networks with a power-law degree
distribution $P(k) \propto k^{-\lambda}$ is needed. Buldyrev et
al. \cite{Sergey2010} developed a framework, based on percolation
theory, to study the robustness of interdependent networks. Analysis
of  {\it fully} interdependent scale-free networks (where all nodes in one
network depend on all nodes in the other network and vice versa)
shows \cite{Sergey2010} that, the
critical threshold is $p_c > 0$ even for $\lambda \leq 3$, in
contrast to a single network where $p_c =0$ \cite{Cohen2000}. In
general, for fully interdependent networks with the same average
degree, the broader the degree distribution is (smaller value of
$\lambda$), $p_c$ is larger \cite{Sergey2010}. This means that networks with a broader
degree distribution become less robust compared to networks with a
narrower degree distribution. This feature is in contrast to the
trend known in single non-interacting networks where networks with broader
degree distribution are more robust. In real world, however, not all
nodes in one network depend on all nodes in the other network, so it
is of interest to study the robustness of two partially interdependent scale-free
networks. Parshani et al. \cite{Rony2010} generalized
the above framework \cite{Sergey2010} to study partially
interdependent networks. Ref.~\cite{Rony2010}
studied the case of partial coupling where only a fraction $q$ of nodes
in each network are interdependent. Their results
for two interdependent Erdos-Renyi (ER) \cite{ER1959, Bollobas1985} networks
show that
there exists a critical $q_c$, bellow which the system shows a
second order percolation transition while above $q_c$ a first order discontinuous
percolation transition occurs. The evolution of such
a change from first order to second order for SF networks when $q$
changes remained unclear, because the behavior of interdependent
SF networks is much more complex.

In this paper, we study the robustness of two partially
interdependent scale-free (SF) networks under random attack. We assume that
only a fraction $q$ of nodes in each network are
interdependent.
We find that for SF networks there are three types of behaviors for different $q$.
In addition to first-order transition for large $q$ and second order for small $q$ there
is a mixed first-second order transition in intermediate $q$ values.
Specifically, we find (i) As the coupling
strength between the two networks, $q$, reduces from $1$ to $0$, the
giant component, $P_{\infty}$, of the interdependent networks 
show three different types of transitions with $p$.
For $q_1 < q \leqslant 1$, an abrupt collapse transition
occurs. In the range $ q_2 < q < q_1$,
a hybrid transition which is combined of
both abrupt and continuous transitions appears. For $q < q_2$,
a continuous second-order transition appears. (ii) The
threshold $q_1$ which separates the first-order and the hybrid
transition is equal to $1$ for $\lambda \leqslant 3$ and decreases
with increasing $\lambda$. When $q_2 < q < q_1$, at the
steady state of the cascading failures,
there exists a $p$ value, $p_c^{jump}$, at which
the coupled SF networks will
suffer a substantial damage due to cascading failures but
a very small non-zero mutual giant cluster $P_\infty$
will survive.
For $p < p_c^{jump}$, $P_\infty$ will
continuously approaches to $0$
at $p = p_c = 0$ for $\lambda \leqslant 3$ and at $p = p_c > 0$ for $\lambda>3$.
Thus, the theoretical critical
threshold $p_c=0$
for $\lambda \leqslant 3$ for single networks \cite{Cohen2000} is expected to be valid
also for strictly partially
interdependent networks.
(iii) For $q < q_2$, the percolation
transition becomes a regular second order transition,
where $P_\infty$ continuously decreases to zero with decreasing $p$.

\section{Cascading Failures}
\subsection{Initial failure in one network}
When the system contains interdependent networks, which are several
networks fully or partially coupled with each other, the initial
attack on first network can trigger a systematic cascade of failures
between the networks \cite{Sergey2010}. This can be explained as
follows: suppose we have a system of two interdependent
networks $A$ and $B$. When, at the initial attack,
a fraction $1-p$ of nodes in network $A$
($A-nodes$) are removed since
a fraction $q$ of one to one bidirectional dependency links exists
between $A-nodes$ and
$B-nodes$, so these $B-nodes$ which depend on the removed nodes in $A$
are also removed from the network $B$.
Due to initial removal, network A may breaks into some connected
parts,
which are disconnected between themselves, called
clusters. We assume that only the largest cluster
(known as the giant component) will function and all the other
smaller clusters will become dysfunctional. Then the malfunctioning
of the nodes in the small clusters of network $A$ will cause failures of
their counterparts that depend on them in network
$B$, so network $B$ will also breaks into clusters, and will cause
further fragmentation in network A. This cascade of failures will
keep going on iteratively until no further failures will occur.

To theoretically study the pair of coupled SF networks under random
failures, we apply the framework developed by Parshani $et~al$
\cite{Rony2010} to study the cascading failures of partially interdependent
random networks. Define $p_A$ and $p_B$ as the fraction of nodes
belonging to the giant components of networks $A$ and $B$,
respectively. Define $\psi^{\prime}_n$ and $\phi^{\prime}_n$ as the
fraction of network $A$ nodes ($A-nodes$) and network $B$ nodes
($B-nodes$) remaining, and $\psi_n$ and $\phi_n$ the giant components
of networks $A$ and $B$ respectively after the cascade of failures
stage $n$. Since $\psi^{\prime}_1$ stands for the remaining fraction
of $A-nodes$ after the initial removal, it follows that $\psi^{\prime}_1=p$.
The remaining functional part of network $A$ therefore contains a
fraction $\psi_1=\psi^{\prime}_1p_A(\psi^{\prime}_1)$. Because a
fraction $q$ of nodes from network $B$ depends on nodes from network
$A$, the number of nodes in network $B$ which loses functionality is
$(1-\psi_1)q=q[1-\psi^{\prime}_1p_A(\psi^{\prime}_1)]$. Similarly,
$\phi^{\prime}_1=1-q[1-\psi^{\prime}_1p_A(\psi^{\prime}_1)]$, among
these $B-nodes$, the fraction of nodes in the giant component of
network $B$ is $\phi_1=\phi^{\prime}_1p(\phi^{\prime}_1)$. The
general form of the iterations is

\begin{equation}\label{eq1}
\begin{array}{lcl}
\psi^{\prime}_1=p,~~~~~~~~~~~~~~~~~~~~~~~~~~~~~ \psi_1=\psi^{\prime}_1p_A(\psi^{\prime}_1), & \mbox{} & \\
\phi^{\prime}_1=1-q[1-\psi^{\prime}_1p_A(\psi^{\prime}_1)],~~~~~  \phi_1=\phi^{\prime}_1p_B(\phi^{\prime}_1),& \mbox{} & \\
\psi^{\prime}_2=p[1-q(1-p_B(\phi^{\prime}_1))],~~~~ \psi_2=\psi^{\prime}_2p_A(\psi^{\prime}_2)..., & \mbox{} & \\
\psi^{\prime}_n=p[1-q(1-p_B(\phi^{\prime}_{n-1}))],~ \psi_n=\psi^{\prime}_np_A(\psi^{\prime}_n),& \mbox{} & \\
\phi^{\prime}_n=1-q[1-p_A(\psi^{\prime}_n)p],~~~~~~\phi_n=\phi^{\prime}_np_B(\phi^{\prime}_n).
\end{array}
\end{equation}

At the end stage of the cascade of failures when nodes failure
stops, both networks reach a stable state where no further
cascading failures happen. According to Eq.~(\ref{eq1}), it means

\begin{equation}\label{eq1-1}
\begin{array}{lcl}
\phi^{\prime}_m=\phi^{\prime}_{m+1}, & \mbox{} & \\
\psi^{\prime}_m=\psi^{\prime}_{m+1},& \mbox{}
\end{array}
\end{equation}
when $m \rightarrow \infty$, since eventually the clusters stop
fragmenting.

Let $\psi^{\prime}_m$ be denoted by $x$ and $\phi^{\prime}_m$ by
$y$, so we get $\psi_{\infty}=p_A(x)x$, $\phi_{\infty}=p_B(y)y$.
Applying the previous conditions with the last two equations in
Eq.~(\ref{eq1}), we obtain the set of equations
\begin{equation}
 \left\{
  \begin{array}{l l}
    x=p\{1-q[1-p_B(y)]\}\\
    y=1-q[1-p_A(x)p]. \\
  \end{array} \right.
\label{eq2}
\end{equation}

Eq.~(\ref{eq2}) \cite{Rony2010} can be solved numerically to get the values of
$x$ and $y$ when an analytical solution is not possible. This is the
case for coupled SF networks, since the generating
functions of SF network do not have a convergent analytical form,
and only an infinite series can be obtained.

Next we introduce the mathematical technique of generating functions for SF networks
in order to get the analytical forms of $p_A(x)$ and $p_B(x)$ \cite{Sergey2010, Shao2008,
Rony2010, Newman2002}. The generating function of the degree
distribution is

\begin{equation}\label{eq2-1}
G_{A}(z_A)=\sum_{k}P_{A}(k)z_A^{k}.
\end{equation}

Analogously, the generating function of the underlying branching
processes is

\begin{equation}\label{eq2-1}
H_{A}(z_A)=G^{'}_{A}(z_A)/G^{'}_{A}(1).
\end{equation}

Random removal of a fraction $1-p$ of nodes will change the degree
distribution of the remaining nodes, so the generating function of
the new distribution is equal to the generating function of the
original distribution with the argument equal to $1-p(1-z)$
\cite{Shao2009, Newman2002}. The fraction of nodes in A that
belongs to the giant component after the removal of $1-p$ nodes is

\begin{equation}
p_A(p)=1-G_{A}[1-p(1-f_A)], \label{eq3}
\end{equation}
where $f_A$ is a function of $p$, $f_A\equiv f_A(p)$, which
satisfies the transcendental equation
\begin{equation}
f_A=G_{A}[1-p(1-f_A)]. \label{eq4}
\end{equation}
For SF networks, the degree distribution is $P(k)=ck^{-\lambda}$ where $\lambda$ is the broadness of the distribution
and $k_{min}<k<K$.
In the case of SF networks \cite{Havlin2010},
\begin{equation}
G_{A}(z_A)=\sum\limits_{k=k_{min}}^K[(\frac{k_{min}}{k})^{\lambda-1}-(\frac{k_{min}}{k+1})^{\lambda-1}]z_A^k,
\label{eq5}
\end{equation}
and
\begin{equation}
H_{A}(z)=\frac{\sum\limits_{k=k_{min}}^K
k[(\frac{k_{min}}{k})^{\lambda-1}-(\frac{k_{min}}{k+1})^{\lambda-1}]z_A^{k-1}}
               {\sum\limits_{k=k_{min}}^K k[(\frac{k_{min}}{k})^{\lambda-1}-(\frac{k_{min}}{k+1})^{\lambda-1}]}.
\label{eq6}
\end{equation}

From Eqs. (\ref{eq2})-(\ref{eq6}), we obtain that

\begin{equation}\label{eq6-1}
\begin{array}{lcl}
\phi_{\infty}=\frac{(1-z_A)(1-G_A(z_A))}{1-H_A(z_A)}, & \mbox{} & \\
\psi_{\infty}=\frac{(1-z_B)(1-G_B(z_B))}{1-H_B(z_B)},& \mbox{}
\end{array}
\end{equation}
where $z_A$ and $z_B$ satisfy
\begin{equation}\label{eq6-2}
\begin{array}{lcl}
\frac{(1-z_B)}{1-H_B(z_B)}=1-q[1-p(1-G_A(z_A))], & \mbox{} & \\
\frac{(1-z_A)}{1-H_A(z_A)}=p[1-qG_B(z_B)]. & \mbox{}
\end{array}
\end{equation}

\begin{figure}
  \begin{center}
    \mbox{\subfigure[]{\includegraphics[width=3in]{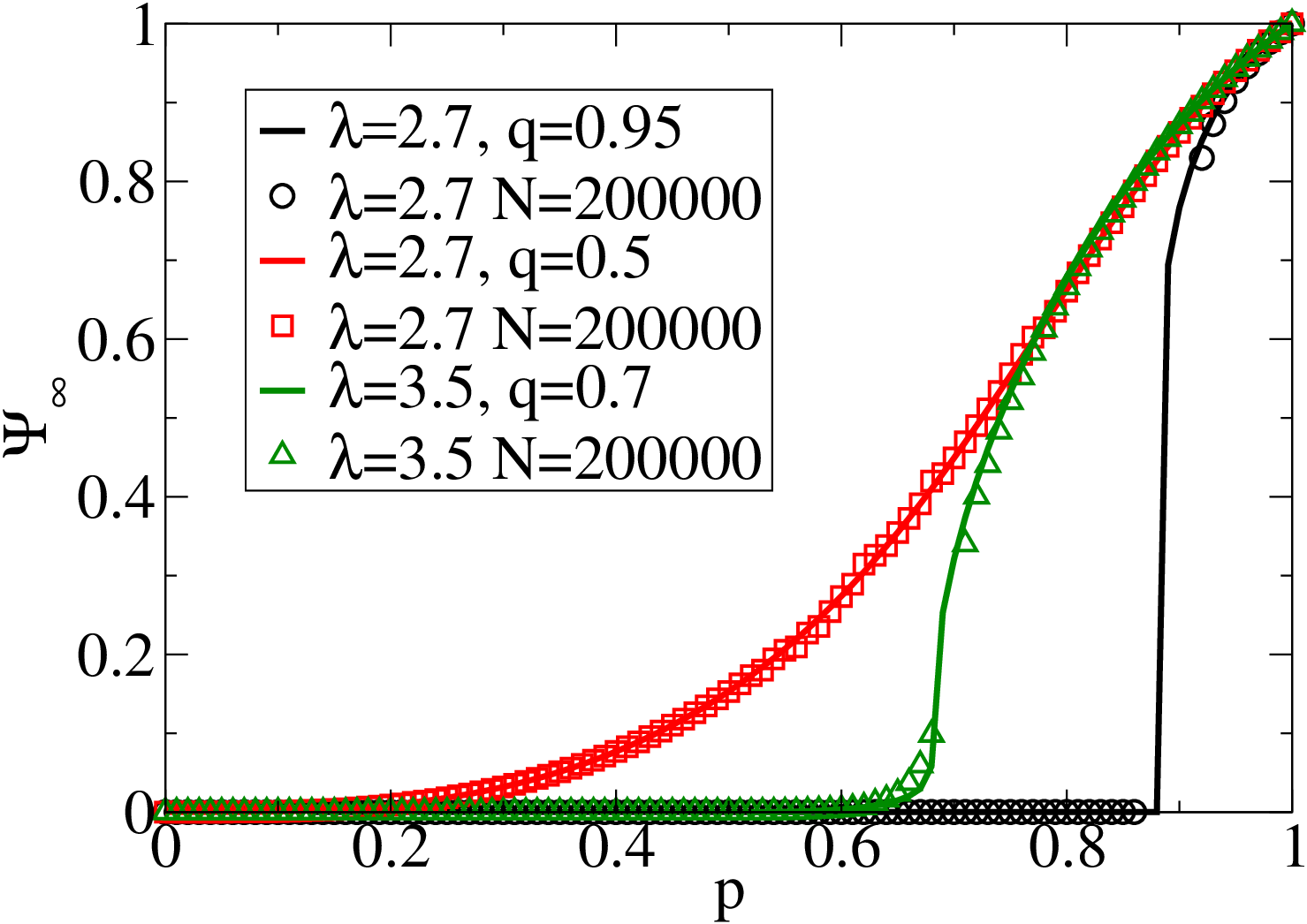}
          \label{Fig1a}}\quad
          \subfigure[]{\includegraphics[width=3in]{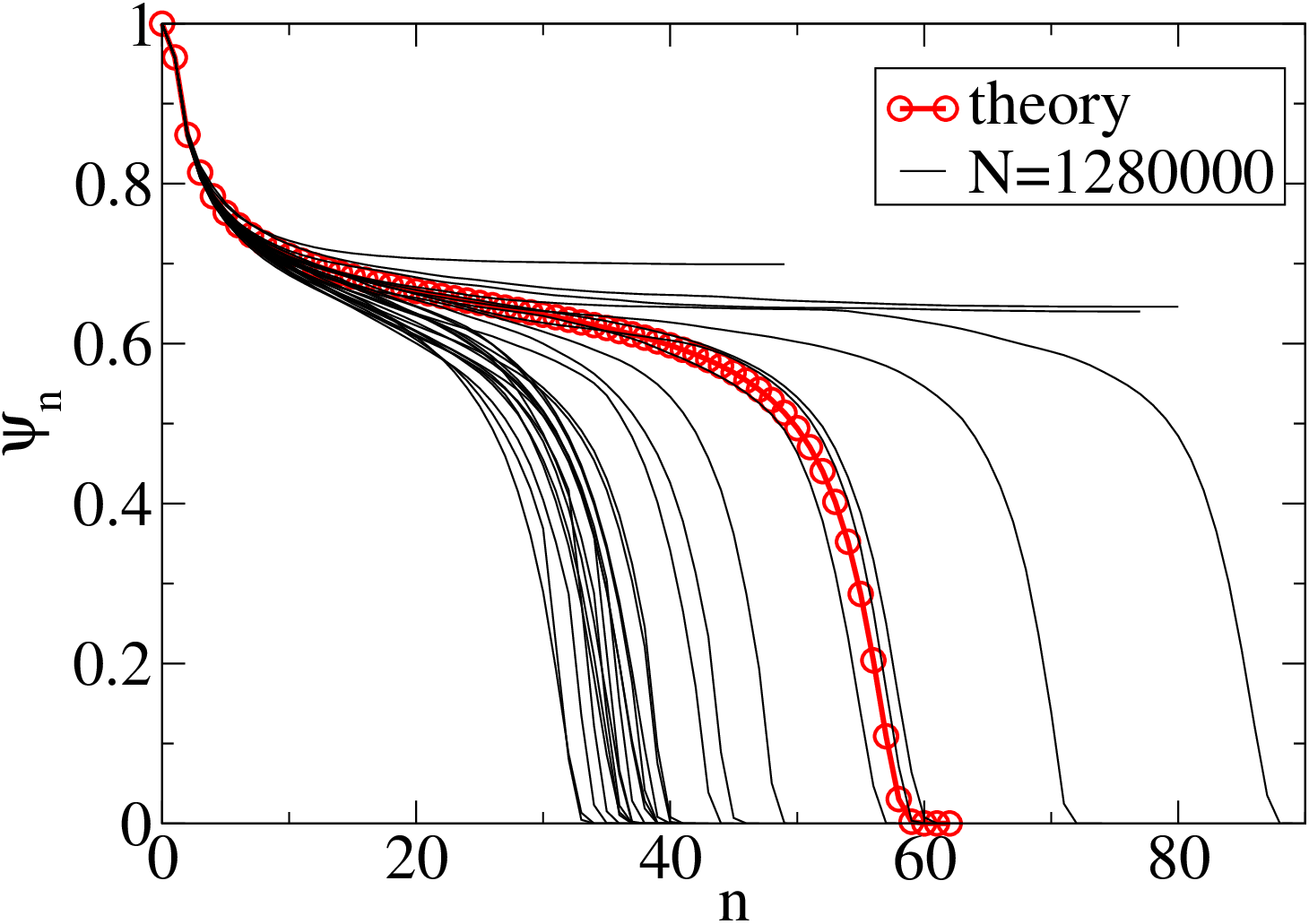}
          \label{Fig1b} }}
  \end{center}
\caption{(Color online) (a) The giant component
$\psi_\infty$ as a function of $p$ for coupled SF-SF networks system under random
removal of $1-p$ nodes in one network.
SF networks with
three different parameters are shown (i) $\lambda=2.7$, $q=0.95$, $k_{min} = 2$, $\langle k \rangle=3$,
(ii) $\lambda=2.7$, $q=0.5$, $k_{min} = 2$, $\langle k \rangle=3$, and
(iii) $\lambda=3.5$, $q=0.7$, $k_{min} = 2$, and $\langle k \rangle=3$.
The lines represent the theory (Eq. (\ref{eq2})) and symbols are results of simulations.
(b) Comparison between theory and simulations of $\psi_n$, the
fraction of the giant component obtained at $p=0.883$, which is just below $p_c$,
after $n$ stages of the
cascading failures for several random realizations of coupled
SF networks with $\lambda = 2.7$, $k_{min} = 2$, $\langle k \rangle=3$, $q = 0.95$, and
$N = 1280000$.
One can see that for the initial stages the agreement is perfect, however
for larger $n$ deviations occur due to random fluctuations in the topology between different realizations
\cite{zhou2012}.
Both simulations and theoretical predictions show a plateau which drops to $zero$,
corresponding to a complete fragmentation of the network.
Note that some of the random realizations
converge to a finite mutual giant component, and are not completely fragmented.
}
\label{Fig1}
\end{figure}

Substituting the generating functions of SF networks into the theoretical
frameworks, Eqs.~(\ref{eq1})-(\ref{eq4}), we obtain, using numerical
solutions, the theoretical results and compare them with results of
computer simulations. Fig.~\ref{Fig1a} shows good agreement between
the theoretical and simulation results for the final giant component
$\psi_\infty$ as a
function of $p$ for two interdependent SF networks
under random removal of $1-p$ nodes in one network.
Three cases are studied: (i) $\lambda=2.7$, $q=0.95$, $k_{min} = 2$, $\langle k \rangle=3$;
(ii) $\lambda=2.7$, $q=0.5$, $k_{min} = 2$, $\langle k \rangle=3$; and
(iii) $\lambda=3.5$, $q=0.7$, $k_{min} = 2$, and $\langle k \rangle=3$.
Fig.~\ref{Fig1b} shows the cascading failure
dynamics of the giant components left after $n$ cascading stages
(denoted by $\psi_n$) as a function of number of iterations $n$, for
several random realizations of SF networks with $\lambda = 2.7$,
$k_{min} = 2$, $\langle k \rangle=3$ (same parameter values as the
numerical calculation), $N = 1,280,000$ at $p = 0.883<p_c$, in
comparison with the theoretical prediction of Eq.~(\ref{eq1}).
Initially the agreement is perfect and when $n$ is getting larger,
the random fluctuations in topology of different realizations play an
important role \cite{zhou2012}.

\subsection{Initial failures in both networks}\label{bnf}
When initially a $1-p$ fraction of nodes is removed from {\it both} networks
\cite{gao2010,gao2012}, the system equations (\ref{eq2}) becomes

\begin{equation}
 \left\{
  \begin{array}{l l}
    x=p\{1-q[1-p_B(y)p]\},\\
    y=p\{1-q[1-p_A(x)p]\}. \\
  \end{array} \right.
\label{eq6-3}
\end{equation}

When the degree distribution of the two networks are the
same, it follows that $p_B(\cdot)=p_A(\cdot)$, $x=y$ and $\phi_{\infty} =
\psi_{\infty}$, and the two equations (\ref{eq6-3}) become a single
equation. Furthermore, using Eqs. (\ref{eq6-1}) and (\ref{eq6-2}),
we obtain
\begin{equation}\label{eq6-4}
\psi_{\infty}=\phi_{\infty}=\frac{(1-z)(1-G(z))}{1-H(z)},
\end{equation}
where $z$ satisfies
\begin{equation}\label{eq6-5}
\frac{(1-z)}{1-H(z)}=p\{1-q[1-p(1-G(z))]\}.
\end{equation}

Eq. (\ref{eq6-5}) is a quadratic equation of $q$, and only
one root has a physical meaning as

\begin{equation}\label{eq6-6}
\frac{1}{p}=\frac{(H(z)-1)[1-q+\sqrt{((1-q)^2+4q\phi_{\infty}(z))}]}{2(z-1)}\equiv
R(z).
\end{equation}
The maximum of $R(z)$ corresponds to $p_c$, and

\begin{equation}\label{eq6-7}
p_c = \frac{1}{\max\{R(z_c)\}},
\end{equation}
where $z_c$ is obtained when $z\rightarrow 1$, i.e.,
$\phi_{\infty}=0$, and thus
\begin{equation}\label{eq6-8}
\max\{R\}=\lim_{z\rightarrow 1}\frac{H(z)-1}{z-1}(1-q)\doteq
H^{\prime}(1).
\end{equation}
For two interdependent SF networks, when $K\rightarrow \infty$,
$\max\{R\}\rightarrow \infty$, so $p_c=0$. However, in the numerical
simulations, $K$ can not reach $\infty$, so $p_c$ seems greater than
0, but in the theory $p_c=0$. Note that when $q=1$, Eq. (\ref{eq6-8})
can yield for $\max\{R\}$ a finite value since $1-q = 0$ and 
therefore $p_c$ can become different from zero as found earlier.

Now let us relate $p_c$ of failure in one network
($p^o_c$) and both networks failures ($p^b_c$). Our previous results
\cite{gao2011pre} show that for two networks $(p^o_c)^2=p^b_c$,
so we argue that for two SF networks, when $p^b_c=0$,
it follows that $p^o_c=0$.

\section{Percolation behavior}

It is known that due to the existence of the interdependence links,
when the two-network-system is under random attack, the iterative
cascade of failures in both networks may result in a percolation
phase transition that completely fragments both networks when the
initial fraction of removed nodes is above the critical threshold.
When all nodes in both networks have 1-on-1 dependency links towards
their counterpart nodes in the other network (given the size of both
networks is the same), i.e., $q=1$, the percolation phase
transition is discontinuous and first order \cite{Sergey2010}; and
when the coupling strength $q$ reduces to $0$ (which becomes the case of a
single SF network), a second order percolation transition exists
\cite{Cohen2000}. However, the change of transition from
first to second order for SF networks when $q$ changes remained
unclear. For coupled Erdos-Renyi (ER) \cite{ER1959, Bollobas1985}
networks having Poissonian degree distribution a critical point
$q_c$ exists. For $q>q_c$ a first order transition occurs while for
$q<q_c$ a second order continuous phase transition occurs
\cite{Rony2010}.

\begin{figure}

    \vspace{-1cm}
    \subfigure
    {
        \includegraphics[width=2.3in]{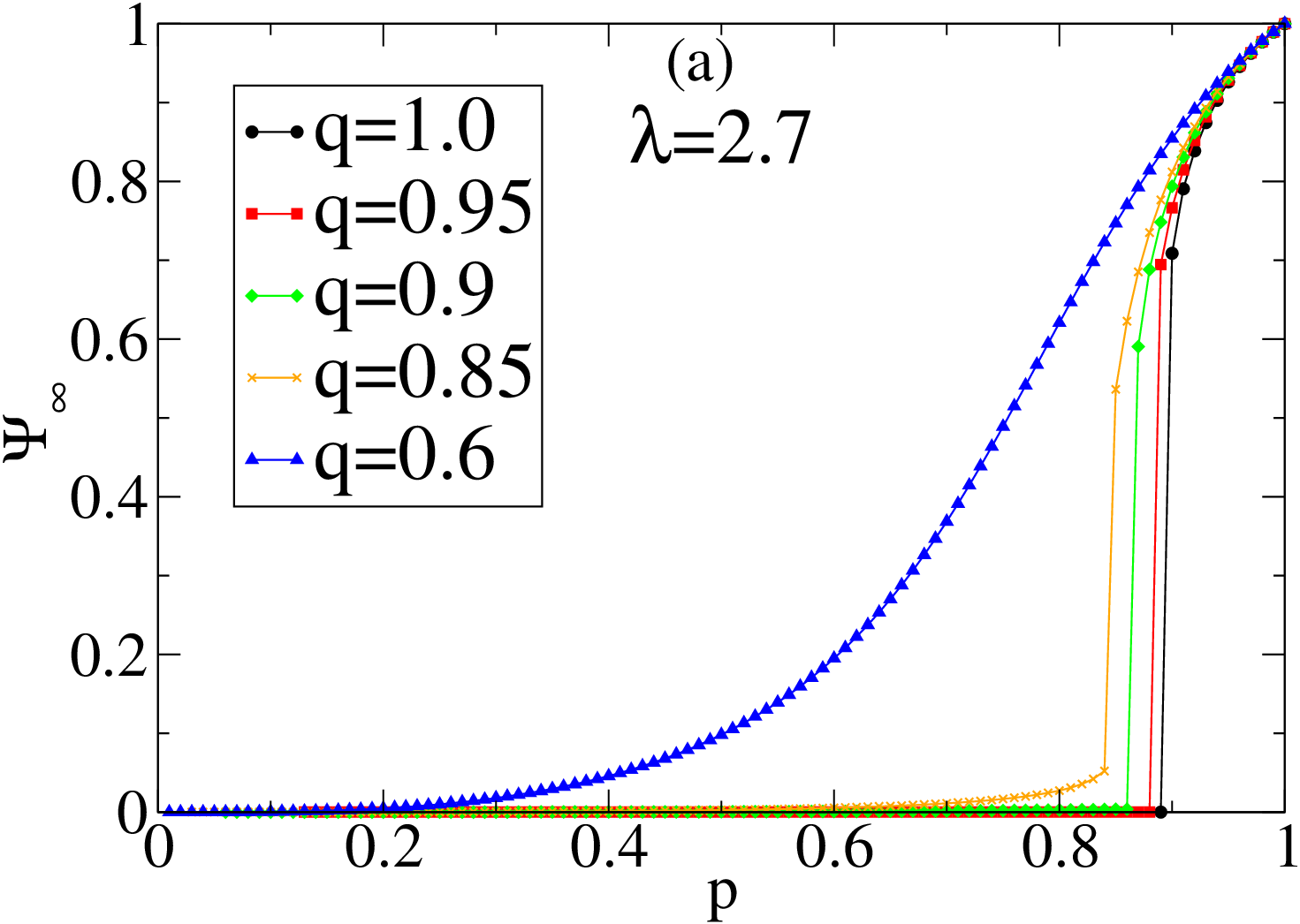}
        \label{Fig2a}
    }
   \vspace{-0.8cm}
    \subfigure
    {
        \includegraphics[width=2.3in]{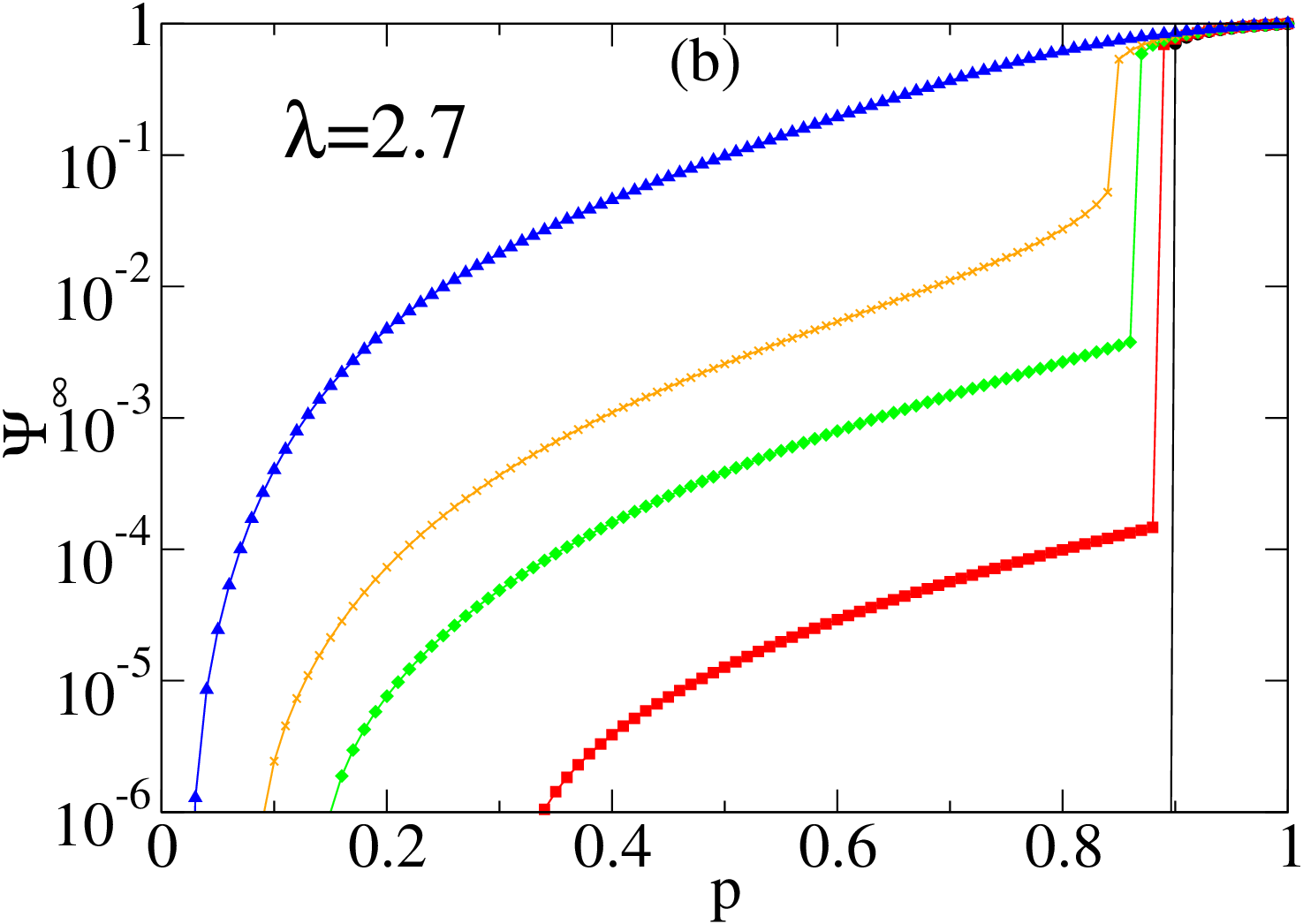}
        \label{Fig2b}
    }
    \vspace{-0.8cm}
    \newline
    \newline
    \newline
    
    \subfigure
    {
        \includegraphics[width=2.3in]{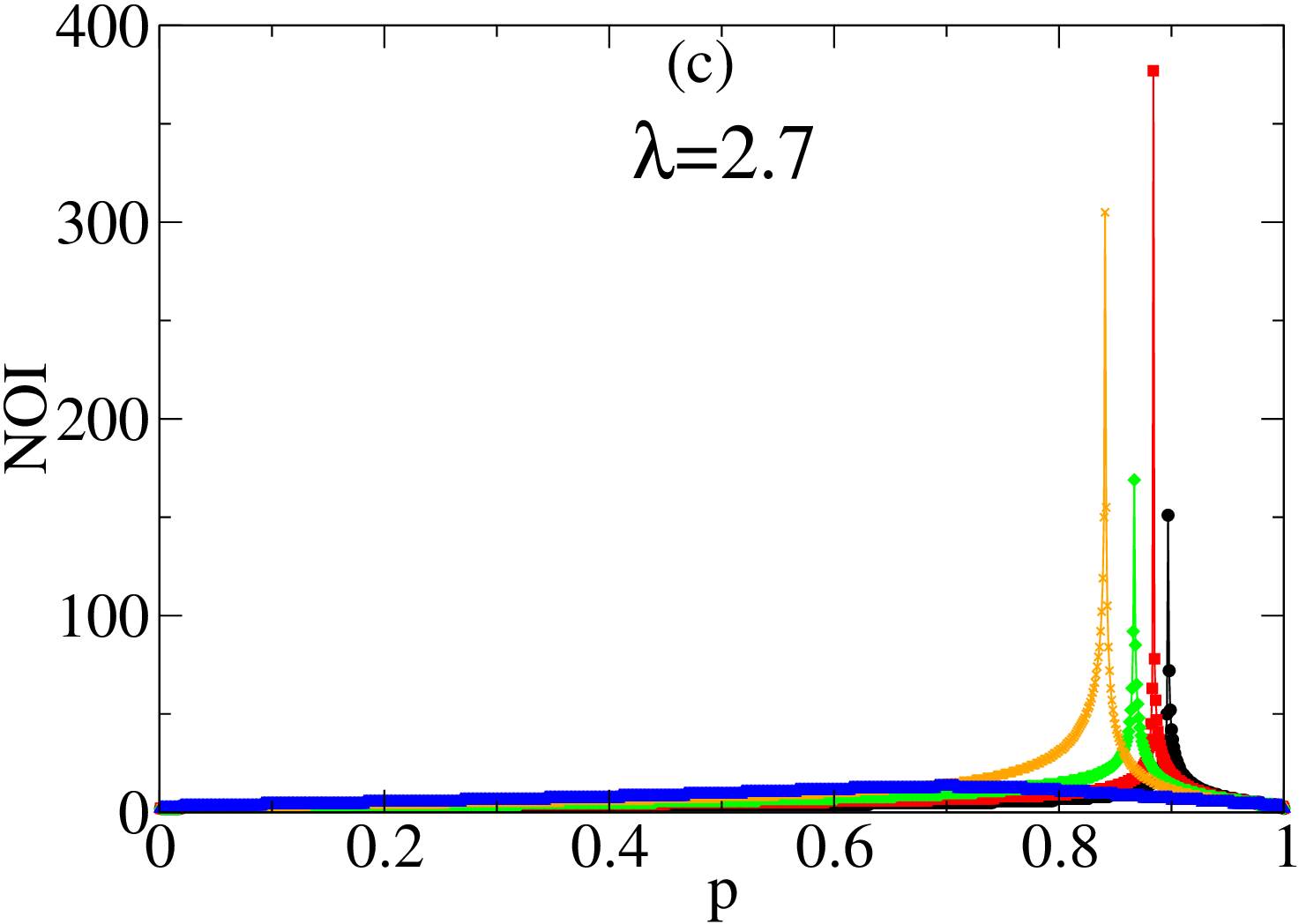}
        \label{Fig2c}
    }
    \subfigure
    {
        \includegraphics[width=2.3in]{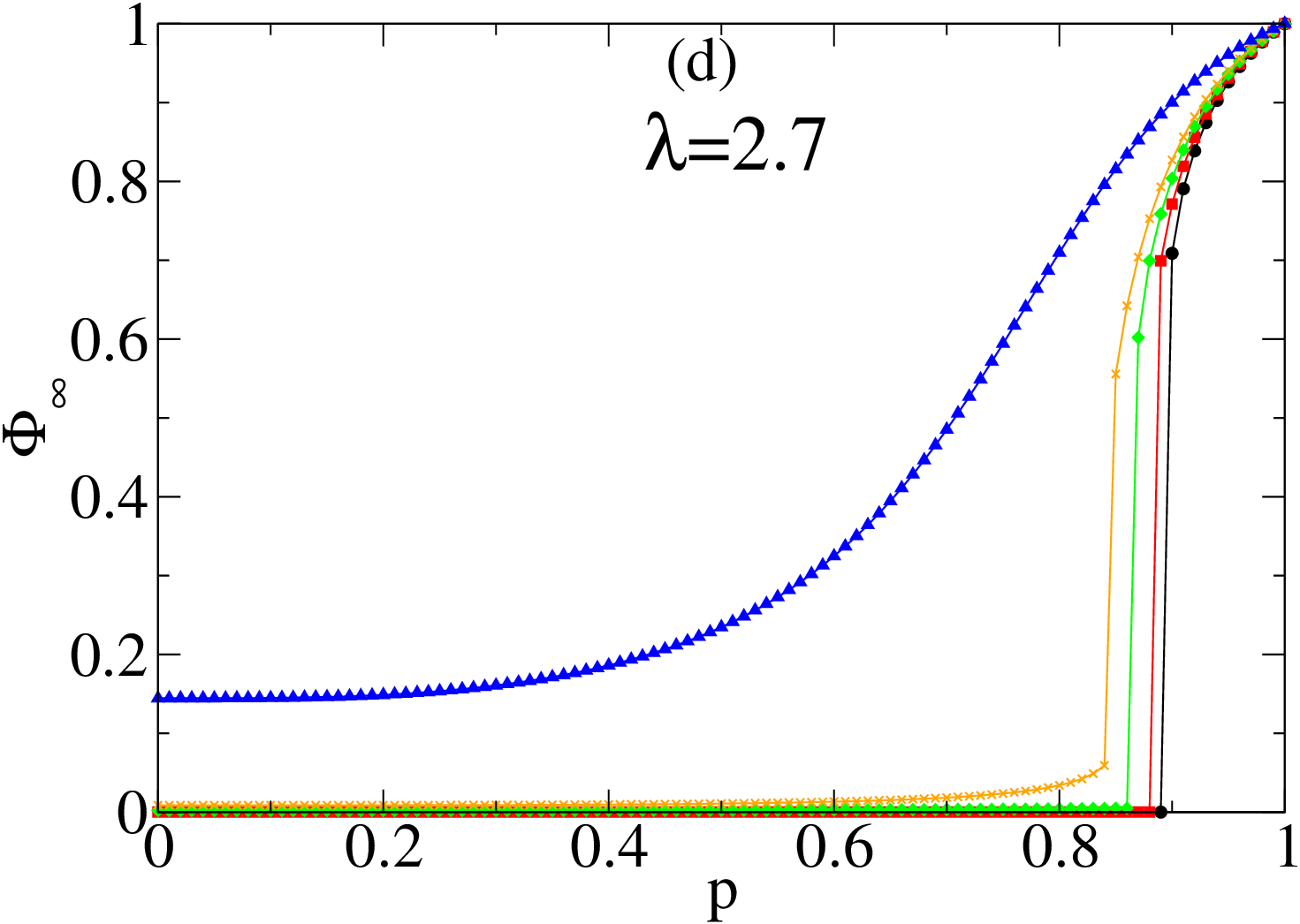}
        \label{Fig2d}
    }
   \vspace{-0.8cm}
    \newline
    \newline
    \newline
    \subfigure
    {
        \includegraphics[width=2.3in]{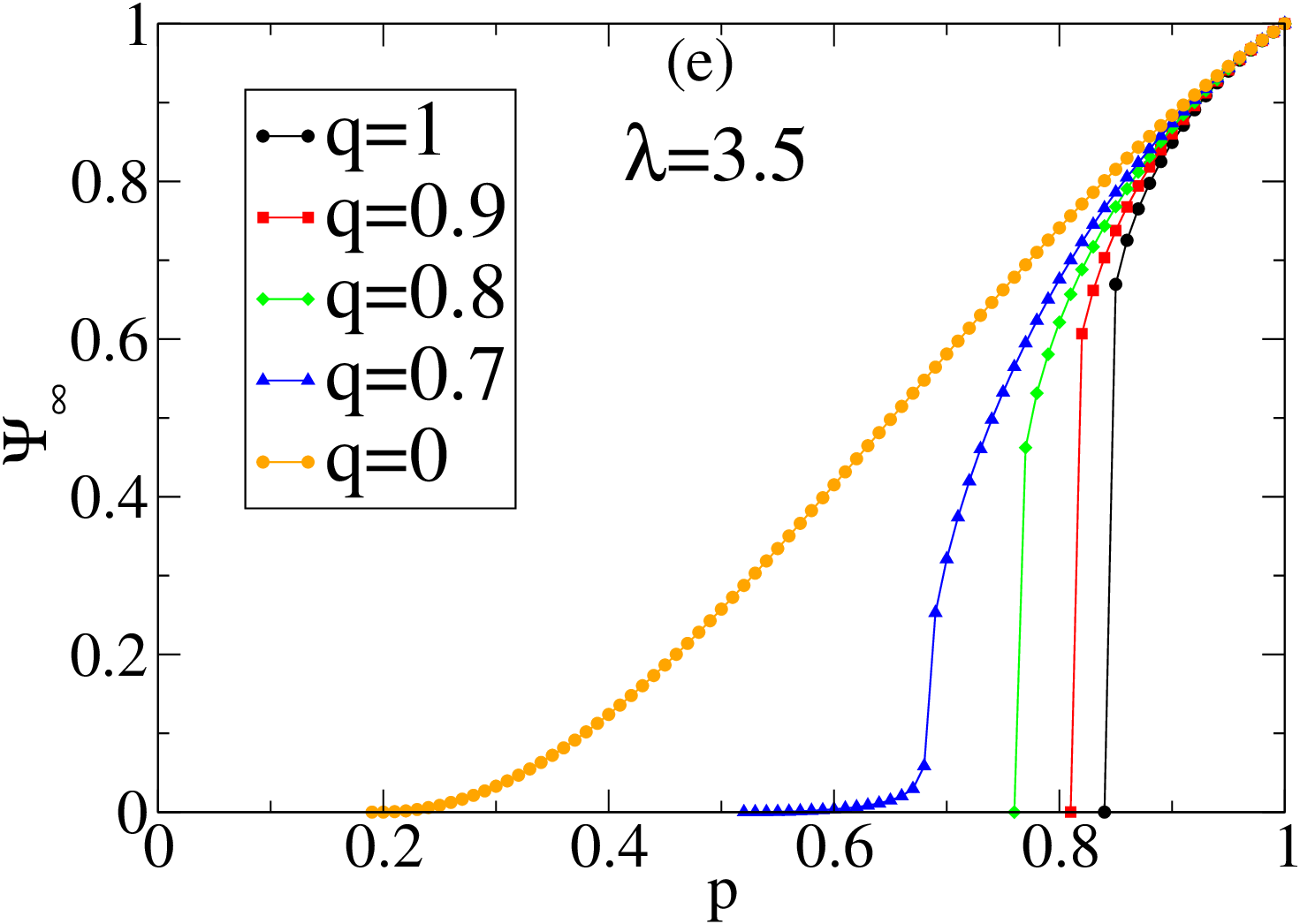}
        \label{Fig2e}
    }
    \vspace{-0.8cm}
    \subfigure
    {
        \includegraphics[width=2.3in]{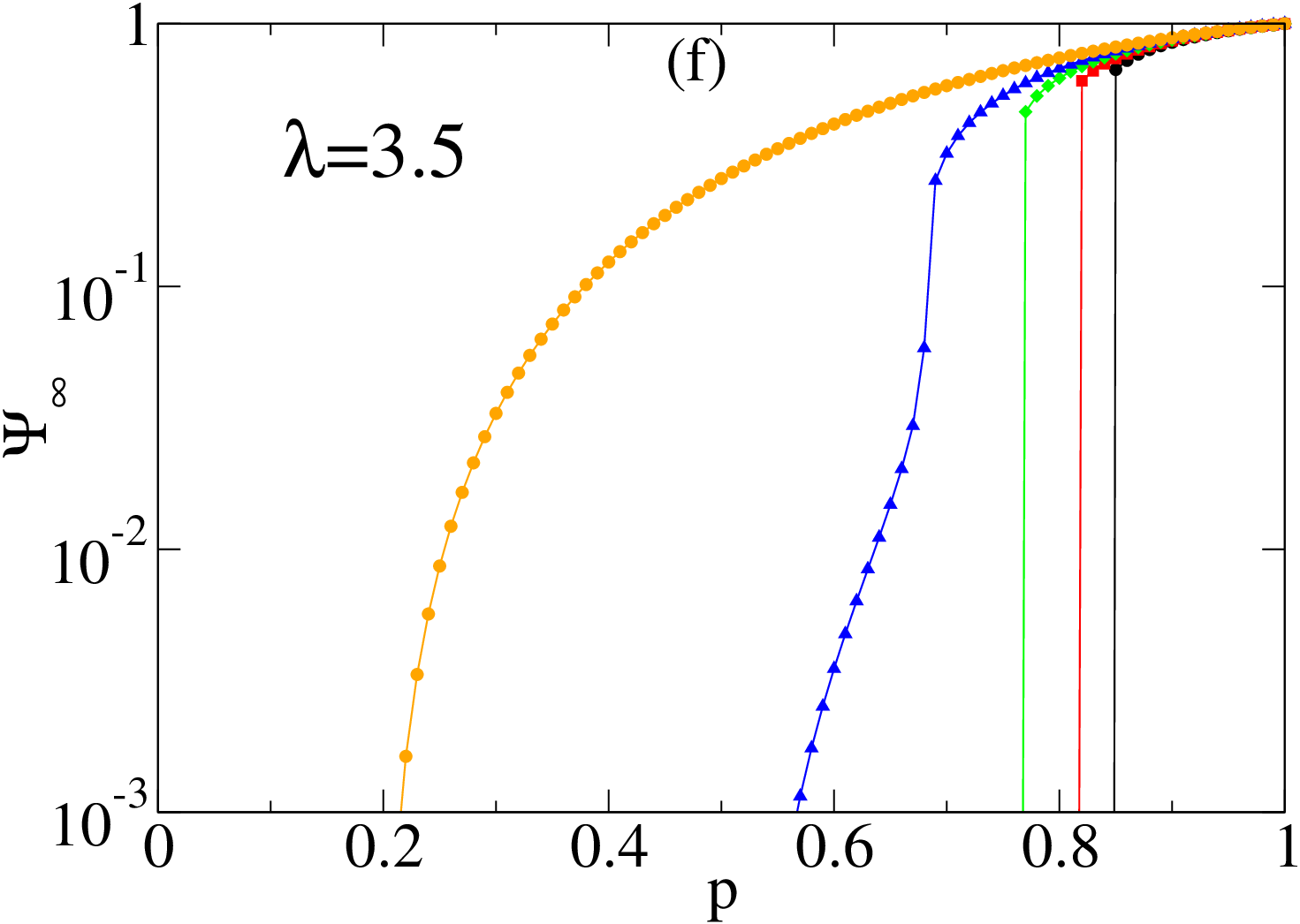}
        \label{Fig2f}
    }
    \hspace{1.2cm}
    \vspace{1cm}

\caption{(Color online) (a) and (b) Numerical calculations of coupled SF networks with
$\lambda = 2.7$, $k_{min} = 2$, average degree $\langle k \rangle = \langle k_A \rangle = \langle k_B \rangle = 3$.
The size of the
giant mutually connected component, $\psi_\infty$, is shown as a function of $p$ for several different values of $q$.
Note that (b) is the same as (a) only that the y-axis is in a logarithmic scale.
We see that when $q=1$, it is a first order transition since $\psi_\infty$ goes to zero for $p$ below the jump
($p_c^{jump}$),
but for $q=0.95$, $q=0.9$, and $q=0.85$,
just below $p_c^{jump}$,
$\psi_\infty$ first reaches a small $non-zero$ value, then smoothly
goes to $zero$ at $p=p_c=0$ (For analytical proof, see Section \ref{bnf}). 
This is a typical property of hybrid phase transition.
For $q=0.6$ there seems to be no jump of $\psi_\infty$
and the transition is purely second order.
(c) The Number-of-Iterations (NOI) \cite{Cohen2011} to reach the end stage of cascade failure
as a function of $p$.
(d) Same plot as (a),
but for $\phi_\infty$, which is the giant component of network B,
which is not initially attacked. 
(e) and (f) are the same as (a) and (b) but for $\lambda = 3.5$, and for $q=1,~0.9,~0.8,~0.7$ and $0$.
}
    \label{Fig2}
\end{figure}

The percolation behavior of two fully and partially
interdependent SF networks, obtained from the numerical solutions of
Eqs.~(\ref{eq2}-\ref{eq6}), are shown in Fig. 2. Figs.~\ref{Fig2a} and (b) show
for $\lambda = 2.7$, the
fraction of nodes in the giant component of network A,
$\psi_{\infty}$, as a function of $p$ (fraction of the initially
unremoved nodes) for several $q$ values. We can see, as expected, for
SF networks, when $q=1$ (fully coupled), the phase transition is
first order \cite{Sergey2010}. This means as more and more nodes are
initially removed, abruptly, at some value of $p=p_c$, the critical
threshold, the iterative cascading failure process will completely
fragment the system. Below $p_c$, there will not exist any cluster
of the order of the network size. Thus, what still will remain are
only very small clusters or single nodes. But just above this
critical $p$ value, when the failures stops, there exists a giant
component in the system.

When $q<1$ but close to 1  ($\lambda \leqslant 3$), as $p$ decreases from $1$,
$\psi_{\infty}$ first shows a sudden big drop similar to $q=1$ case,
but $\psi_{\infty}$ does not drop to $0$, instead, it drops to a small but still
$non-zero$ value, which means though the giant cluster in the
network suffers a big damage, it does not collapse completely (see Fig.~\ref{Fig2b}).
We name the
$p$ value where $\psi_{\infty}$ has the discontinuous drop to be
$p^{jump}_c$. We mathematically
define the $p^{jump}_c$ as

\begin{equation}\label{eq7}
p^{jump}_c=\big
\{p~|~\max\{\psi_{\infty}(^+p)-\psi_{\infty}(^-p)\}\big \},
\end{equation}
where $^+p$ denotes approaching $p$ from above $p$,
and $^-p$ denotes approaching $p$ from bellow $p$.

As $p$ keeps decreasing below $p^{jump}_c$, the small giant
component, $\psi_{\infty}$, smoothly decreases, until at $p = p_c = 0$,
$\psi_{\infty}$ will also reach $0$. Thus, the real critical
threshold for $q<1$ is $p_c=0$ similar to single networks
\cite{Cohen2000} 
(see the analytical arguments at the end of Section \ref{bnf}). 
This phenomenon can be seen more clearly in
Fig.~\ref{Fig2b}, which is similar to Fig.~\ref{Fig2a} but the
y-axis, $\psi_{\infty}$, is plotted in a logarithmic scale. We see
that at $p^{jump}_c$, for $q=0.95$, $q=0.9$, and $q=0.85$, the
corresponding giant component sizes are reduced from order of $1$ by a factor in the range of

\begin{equation}\label{eq8}
\left\{
\begin{array}{lcl}
\psi_{\infty}(^-p^{jump}_c) \in [10^{-2},10^{-4}], & \mbox{} & \\
\psi_{\infty}(^+p^{jump}_c) \approx o(1). & \mbox{}
\end{array}\right.
\end{equation}

When $p$ decreases further, $\psi_{\infty}$ decreases smoothly
towards zero for $p = 0$ (The analytical proof is given in Section \ref{bnf}).

This behavior is typical of the behavior of a
hybrid-transition, which includes both first and second order
phase transition properties similar to that found in bootstrap 
percolation \cite{Dog2010,Schneider2010,Hu2011}.
The giant
component first undergoes a sharp jump, which is a characteristic of
first-order transition, and then smoothly goes to $0$, which is a
characteristic of a second-order phase transition. However, when $q$
is getting smaller, this hybrid-transition phenomenon becomes less
apparent, and the percolation phase behavior seems to become, at some threshold of $q = q_2$,
an ordinary second-order transition. For
example, the curve for $q=0.6$ in Fig.~\ref{Fig2a} and Fig.~\ref{Fig2b}
seems to suggest a second-order transition since there is no obvious
sudden drop of the giant component size, instead, it continuously
decreases when $p$ decreases.
For the case of two interdependent ER networks, the
system shows either a first order
or second order phase transition but not a hybrid transition as here \cite{Sergey2010,gao2010,gao2011pre,gao2012}.

In network $B$, which is initially not attacked,
similar behavior of the giant component $\phi_{\infty}$ can be observed,
see Fig.~\ref{Fig2d}. However, the difference is that even at $p=0$,
$\phi_{\infty}$ does not approach to $0$, but
reaches a finite value. This can be understood due to the
partial dependency between the networks ($q<1$). Even if all nodes in A are removed
($p=0$), since $q<1$, there is a finite fraction, $1-q$, of nodes
in B that are not removed and in a SF network any finite fraction of
unremoved nodes will yield a giant component \cite{Cohen2000}. Only
in the fully coupled ($q=1$) case, the mutually connected giant
cluster will completely collapse at $p_c>0$.

\subsection{Estimate of $p^{jump}_c$ from $P_{\infty}$ as a function of $p$}
\begin{figure}
    \centering
    \subfigure
    {
        \includegraphics[width=3in]{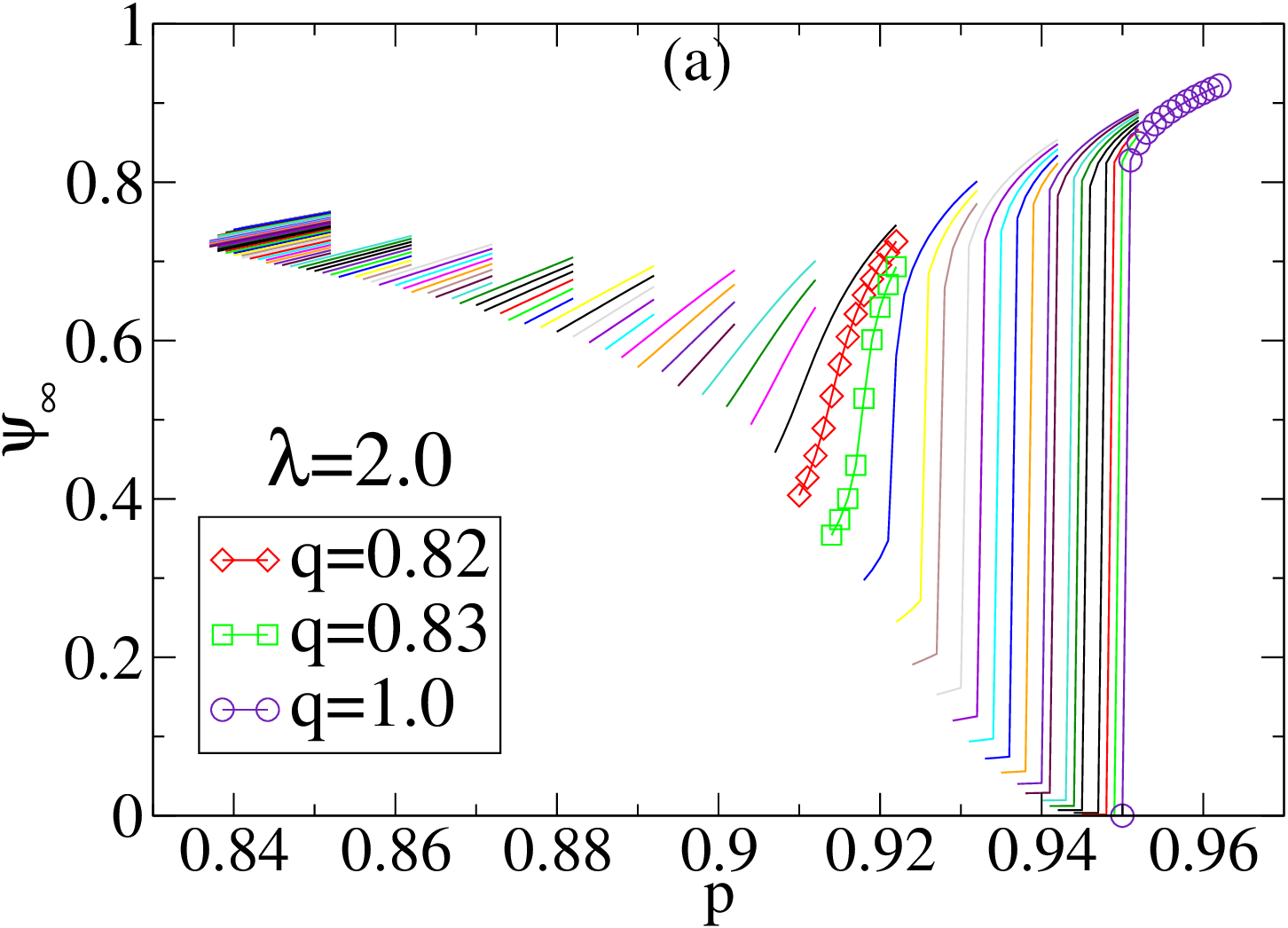}
        \label{Fig3a}
    }
    \subfigure
    {
        \includegraphics[width=3in]{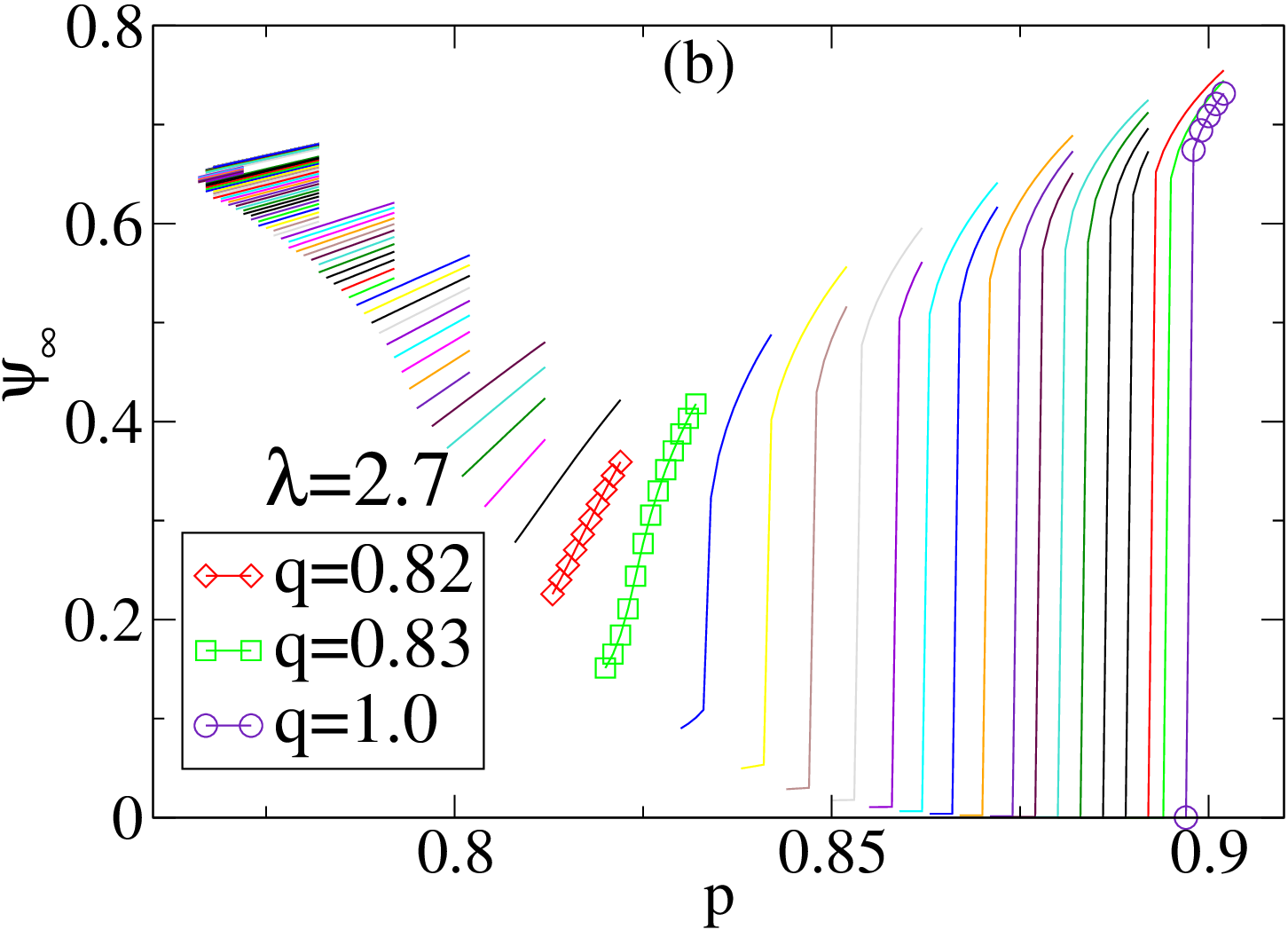}
        \label{Fig3b}
    }
    \newline
    \newline
    \subfigure
    {
        \includegraphics[width=3in]{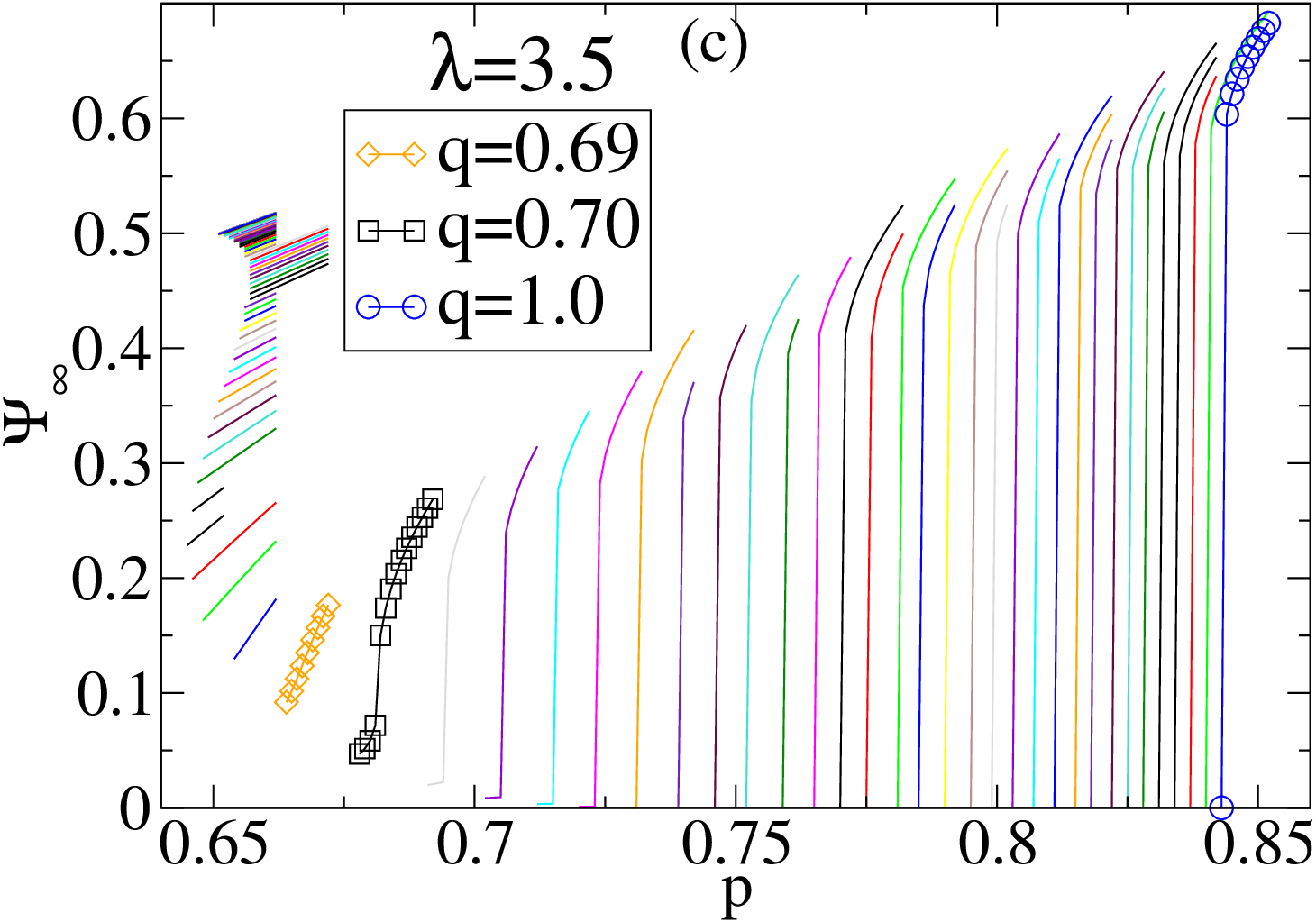}
        \label{Fig3c}
    }
    \caption{ (Color onine) The giant component, $\psi_\infty$ as a function of $p$ for coupled SF networks with
    different values of $\lambda$, with $k_{min} = 2$,
    and average degree $\langle k \rangle = 3$. Only the critical region around
    the maximum jump of $\psi_\infty$ are shown, for different $q$ values ranging from $0$ (most left)
    to $1$ (most right), with increments of $q$ of $0.01$.
    From these graphs, we can
    find as the $q$ decreases, $\psi_\infty$ becomes more continuous. It is also
    seen that for large $q$ the sharp jump of $\psi_\infty$ starts from small but $non-zero$
    values to large finite values. This behavior is typical to a hybrid phase transition.
    (a) $\lambda=2.0$, the threshold of hybrid transition and second-order transition is $q_2 \cong 0.825$,
     so the $q=0.82$ and $q=0.83$ curves are shown with symbols. We can see that the jump in $\psi_\infty$ vanishes
     (shown by symbols) when $q$ is reduced,
     as the phase transition becomes second-order. (b) For $\lambda=2.7$ and (c) for $\lambda=3.5$,
     the curves in the region of $q$ where the hybrid transition becomes second-order are shown by symbols.
     }
    \label{Fig3}
\end{figure}

So far we saw (Fig.~\ref{Fig2}) that for $q_2 < q < q_1$, as $p$ decreases, the
giant component shows an abrupt drop similar to
a first order transition as Eq.~(\ref{eq8}). However,
the drop is not to $P_\infty = 0$ like in a first order transition 
but to a small finite $P_{\infty}$ value.
As $q$ decreases, as seen in Fig.~\ref{Fig3}, this drop becomes less
sharp and smoother, and tends towards a continuous second order
transition as in Eq.~(\ref{eq9}). We analyzed this transition and find
that the phase transition is 
like a first order transition with a
sharp drop of $P_{\infty}$ at $p^{jump}_c$.  For $q < q_2$, the
hybrid transition diminishes and the behavior becomes a second order
transition with a continuous behavior. We are interested to determine
the values of $q_1$ and $q_2$, which separate the three distinct regions.
In order to achieve that, we first need to
find  $p^{jump}_c$.

To accurately evaluate the values of $p^{jump}_c$ for each $q$, we
compute the number of iterations (NOI) in the cascading
process which shows a maximum at $p_c^{jump}$ \cite{Cohen2011}. 
The NOI is the number of iterative cascade steps it
takes the system to reach the equilibrium stage. In the simulations, NOI=$m$ is
defined by Eq.~(\ref{eq1-1}), i.e., the step where no further damage occurs.
But in the numerical solution,
$\psi_{n}$ is approaching $\psi_{\infty}$ only when $n\rightarrow
\infty$. Here we define NOI=$m$ when

\begin{equation}\label{eq9}
\left\{
\begin{array}{lcl}
\psi_{m}-\psi_{m+1} < \xi, & \mbox{} & \\
\phi_{m}-\phi_{m+1} < \xi, & \mbox{}
\end{array}\right.
\end{equation}
where $\xi$ is a very small number. We choose $\xi=10^{-16}$ in this paper, which
is equivalent to the requirement for the cascading failures to stop in
a two-network system when both have $10^{16}$ nodes.
Note that for other very small values of $\xi$ the position of $p_c^{jump}$ remains the same.

At the first-order and hybrid order transition point, the NOI has
its peak value which drops sharply as the distance from
the transition is increased \cite{Cohen2011}. Thus, plotting the NOI as a function of
$p$ provides a useful and precise method for identifying the
transition point $p^{jump}_c$ of
the hybrid transition. Fig.~\ref{Fig2c} presents such numerical
calculation results of NOI. The transition point, $p^{jump}_c$, can
be easily identified by the sharp peak characterizing the transition
point. According to the NOI, we define $p^{jump}_c$ as

\begin{equation}\label{eq7-2}
p^{jump}_c=\big \{p~|~\max\{\mbox{NOI}(p)\}\big \}.
\end{equation}

From Fig.~\ref{Fig2c}, one can see that the definition of Eq.~
(\ref{eq7}) coincides with the definition of Eq.~(\ref{eq7-2}).

\subsection{Determining $q_2$}
We know that when the transition is second-order, the order
parameter, $\psi_{\infty}$, decreases continuously. As seen above,
$\psi_{\infty}$, has a maximum magnitude change at
$p^{jump}_c$, that can used to identify the position of $p^{jump}_c$ by
Eq.~(\ref{eq7-2}). Thus, we can now investigate these maximum magnitude
changes for different $q$ values at $p^{jump}_c$. In Fig.~\ref{Fig3}
we plot $\psi_\infty$ as a function of $p$ only near
$p^{jump}_c$, for different $q$ values ranging between $0$ and $1$, for several different
$\lambda$ values. In order to estimate when these changes are discrete
and when they are continuous, we define

\begin{equation}\label{eq11}
F(q) \equiv log_{10}(\frac{\psi_{\infty}(^+p^{jump}_c(q))}{\psi_{\infty}(^-p^{jump}_c(q))}).
\end{equation}

The rationale for this is as follows. The quantity $\psi_{\infty}(^+p^{jump}_c)$ is
the value of the order parameter just before the maximum drop at $p > p_c^{jump}$, and
$\psi_{\infty}(^-p^{jump}_c)$ is the value of the order parameter right
after the maximum drop. In hybrid transition, as we discussed,
this change in magnitude is large. However, as $q$ becomes smaller,
and when finally the transition becomes second-order,
the change is continuous and the ratio between the magnitudes in Eq.~(\ref{eq11})
should becomes $1$. Thus, whenever $F(q)$ goes to $0$,
the corresponding $q$ is $q_2$, which is the threshold where the
hybrid-transition turns into a second-order phase transition. When
$F(q)\rightarrow 0$,

\begin{equation}\label{eq12}
\frac{\psi_{\infty}(^+p)-\psi_{\infty}(^-p)}{\mbox{d}p}\|_{p=p^{jump}_c}=0.
\end{equation}

\begin{figure}
\begin{center}
    \subfigure
    {
        \includegraphics[width=0.45 \columnwidth]{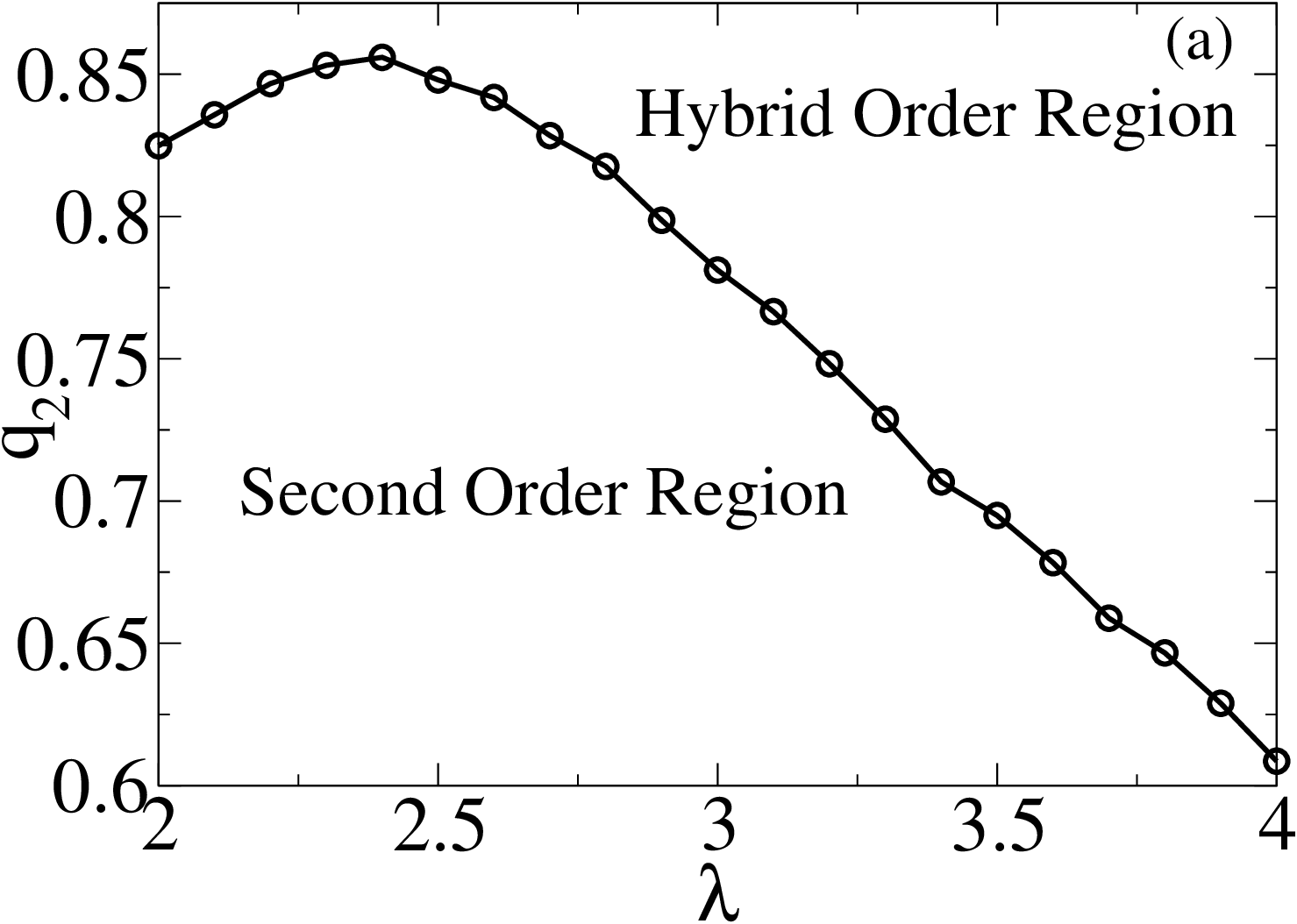}
        \label{Fig4a}
    }
    \subfigure
    {
        \includegraphics[width=0.45 \columnwidth]{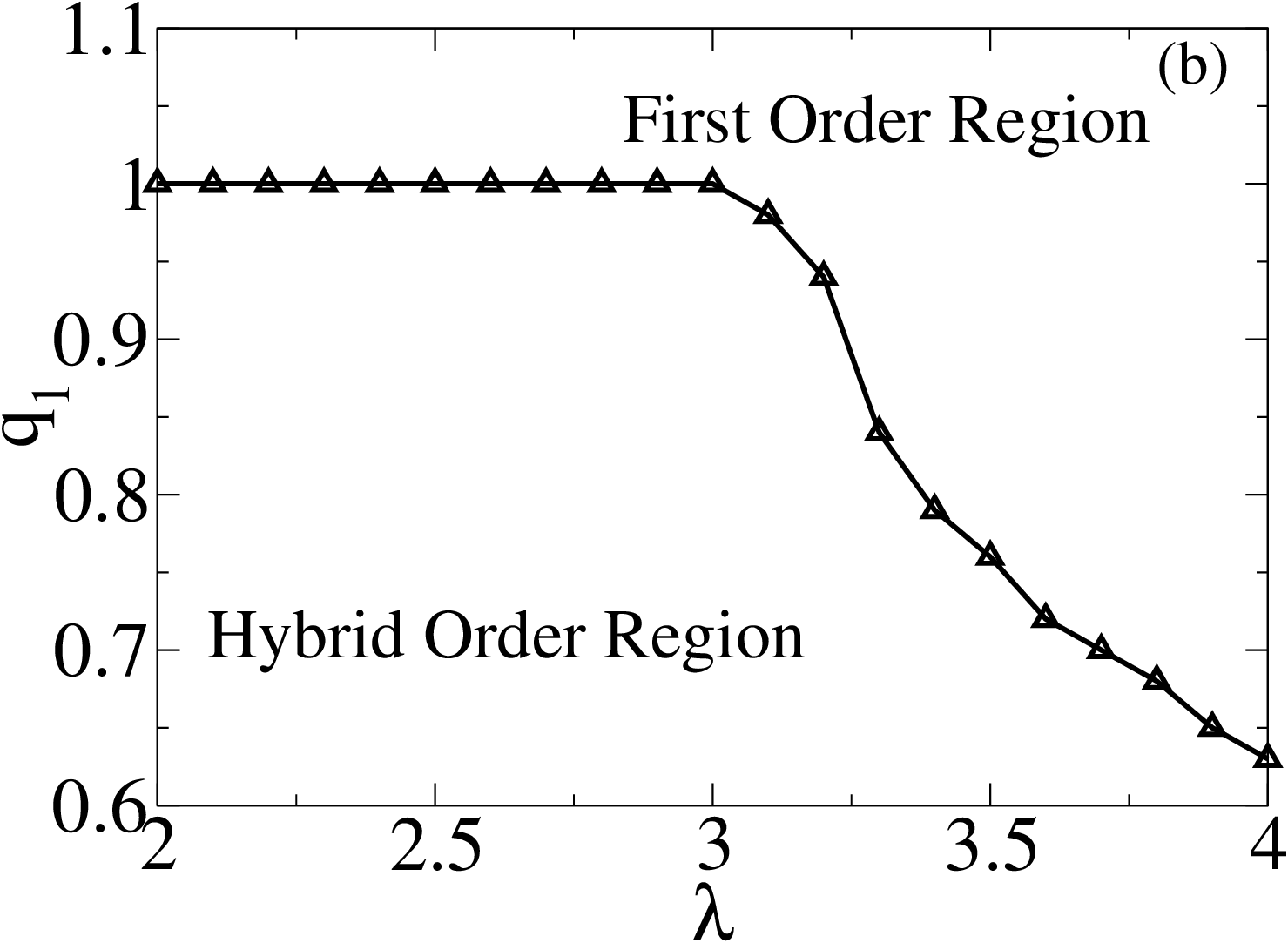}
        \label{Fig4a-2}
    }
    \subfigure
    {
        \includegraphics[width=1.9in]{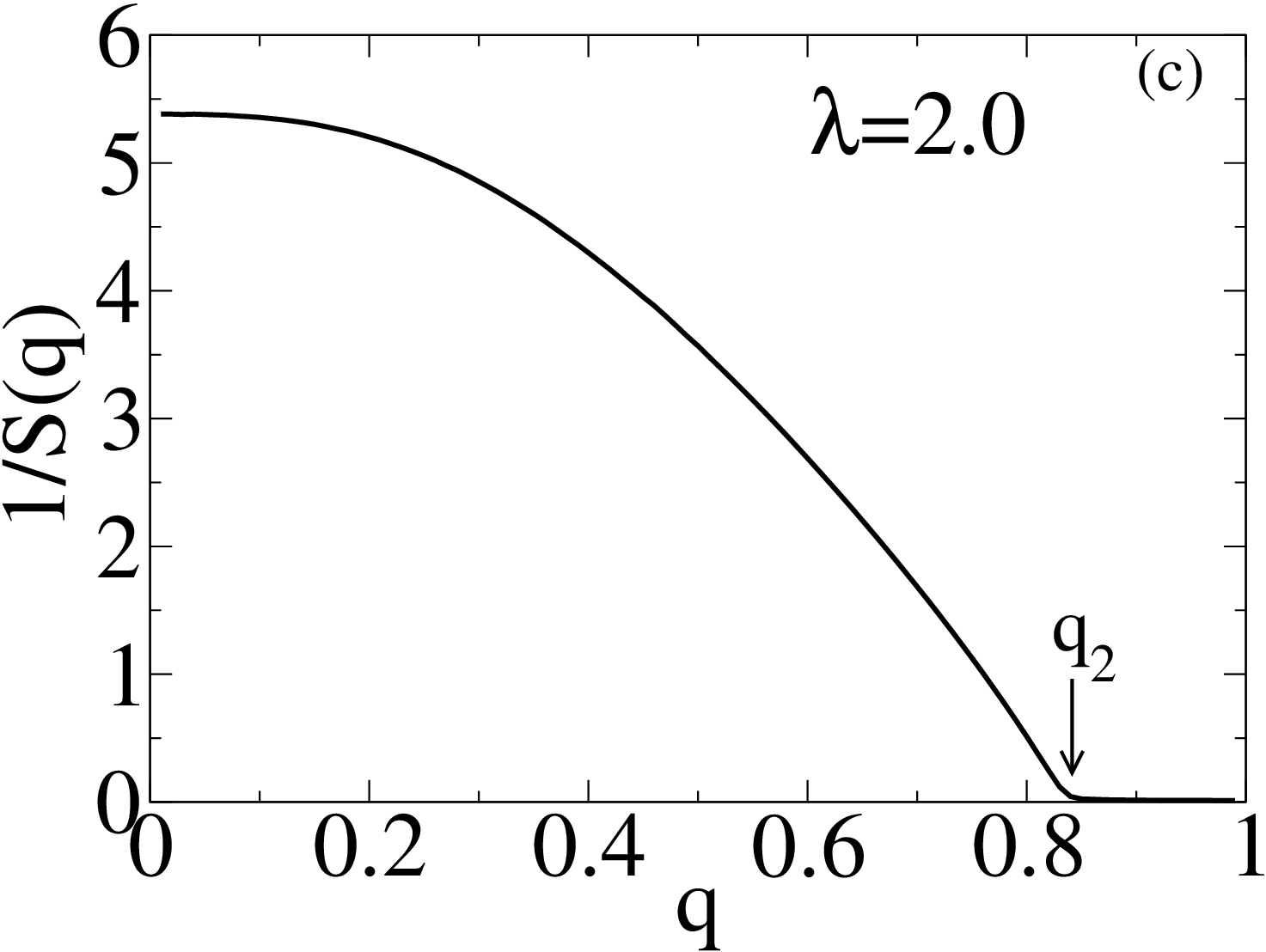}
        \label{Fig4b}
    }
    \subfigure
    {
        \includegraphics[width=1.9in]{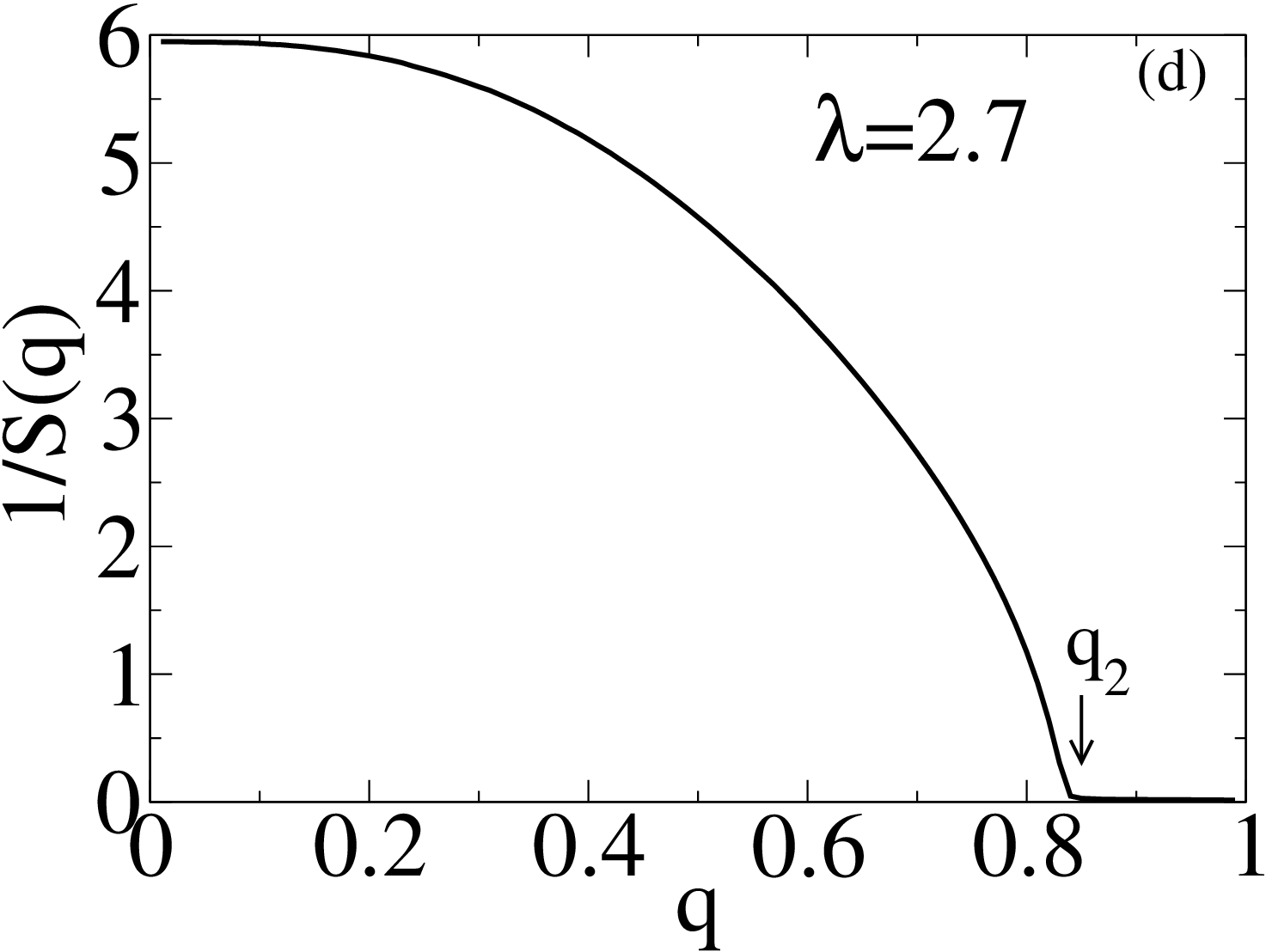}
        \label{Fig4c}
    }
    \subfigure
    {
        \includegraphics[width=2in]{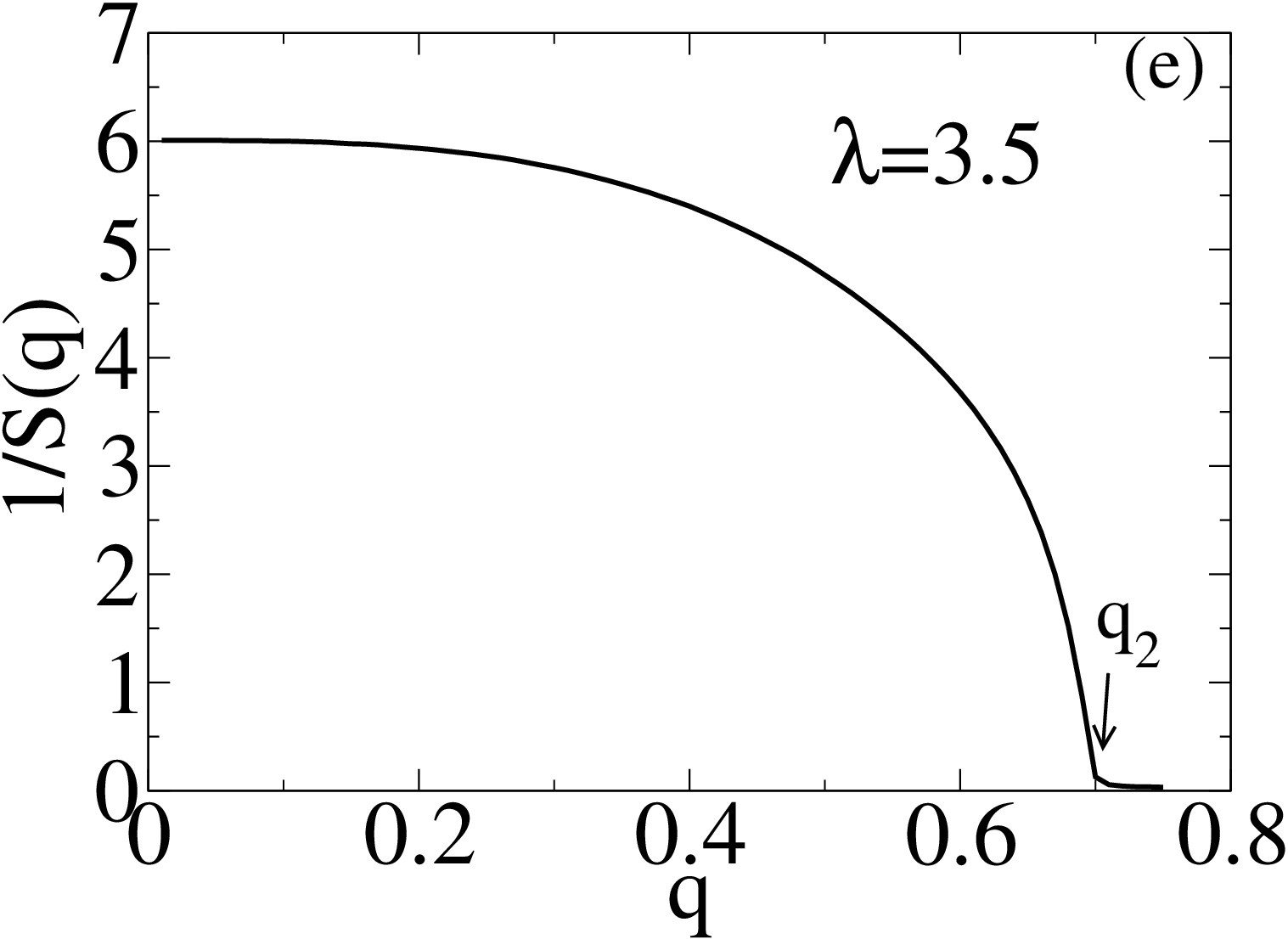}
        \label{Fig4d}
    }
\end{center}
\caption{(Color online) (a) Values of $q_2$ ($\bigcirc$)as a function of $\lambda$ 
 for SF networks with average degree
 $\langle k \rangle=3$ and $k_{min} = 2$.
 Note the maximum of $q_2$ at $\lambda \cong 2.4$.
 (b) Values of $q_1$
 ($\bigtriangleup$) as a function of $\lambda$.
 Plot of $1/S(q)$, as a function of $q$, where $S(q)$ is maximum slope value in the $\psi~vs.~p$ plot,
 are shown for (c) $\lambda=2.0$, (d) $\lambda=2.7$,
 and (e) $\lambda=3.5$, all SF networks are
 with average degree $\langle k \rangle=3$ and minimum degree $k_{min}=2$. We can see that the
 maximum slope values have a sharp change at $q_2=0.83,~0.83,$ and $0.7$ for $\lambda=2.0, 2.7$, and $3.5$
 respectively, supporting the results in (a).
  }
\label{Fig4}
\end{figure}

By extrapolating these $q$ positions (where $F(q)$ goes to $0$) for
different $\lambda$, we get $q_2$ as a function of $\lambda$ and
plot it in Fig.~\ref{Fig4a}.
Interestingly, $q_2$ is not monotonic with $\lambda$ but has a maximum around $\lambda = 2.4$.
To alternatively identify $q_2$, we define the maximum slope as
function of $q$ as $S(q)$,
\begin{equation}\label{eq13}
S(q) \equiv \max\{\frac{\psi_{\infty}(^+p)-\psi_{\infty}(^-p)}{\mbox{d}p}\|_{p=p^{jump}_c}\}.
\end{equation}

When $q$ is below or equal to $q_2$, the value of $S(q)$ is very small, representing a continuous
change of $\psi_{\infty}$, which is second-order; when $q$ reaches
some value, the $S(q)$ has a sudden drop at $q_2$, i.e., the maximum
slope becomes dramatically large, representing a sharp change in
$\psi_{\infty}$, which is a sign of the occurrence of
a hybrid transition.

By identifying the position of $q$ where the abrupt drop is located, we can also
find the thresholds $q_2$ which distinguish second-order and hybrid
transition. The results shown in Fig.~\ref{Fig4b}, Fig.~\ref{Fig4c}, and
Fig.~\ref{Fig4d} match very well the results in Fig.~\ref{Fig4a},
supporting our method for determining $q_2$.

\subsection{Determining $q_1$}

For coupled SF networks with $\lambda \leqslant 3$, only when $q=1$
the transition is a first-order, which means
$q_1 = 1$ for $\lambda \leqslant 3$. As $\lambda$ increases
above $3$, $q_1$ becomes smaller than $1$. To estimate the $q_1$
values for $\lambda > 3$, we define according to Eq.~(\ref{eq8}) the system
to have a first order transition
if $\psi_{\infty}$ satisfies

\begin{equation}\label{eq14}
\psi_{\infty}(^-p^{jump}_c,q) < \sigma.
\end{equation}
Otherwise, it is not a first-order transition.
We set here a value $\sigma = 10^{-11}$ but similar results have been obtained for $\sigma = 10^{-12}$ and $10^{-13}$.
We plot $q_1$ as a
function of $\lambda$ obtained this way in Fig.~\ref{Fig4a-2}.

\begin{figure}
\begin{center}
\includegraphics[width=0.8 \columnwidth]{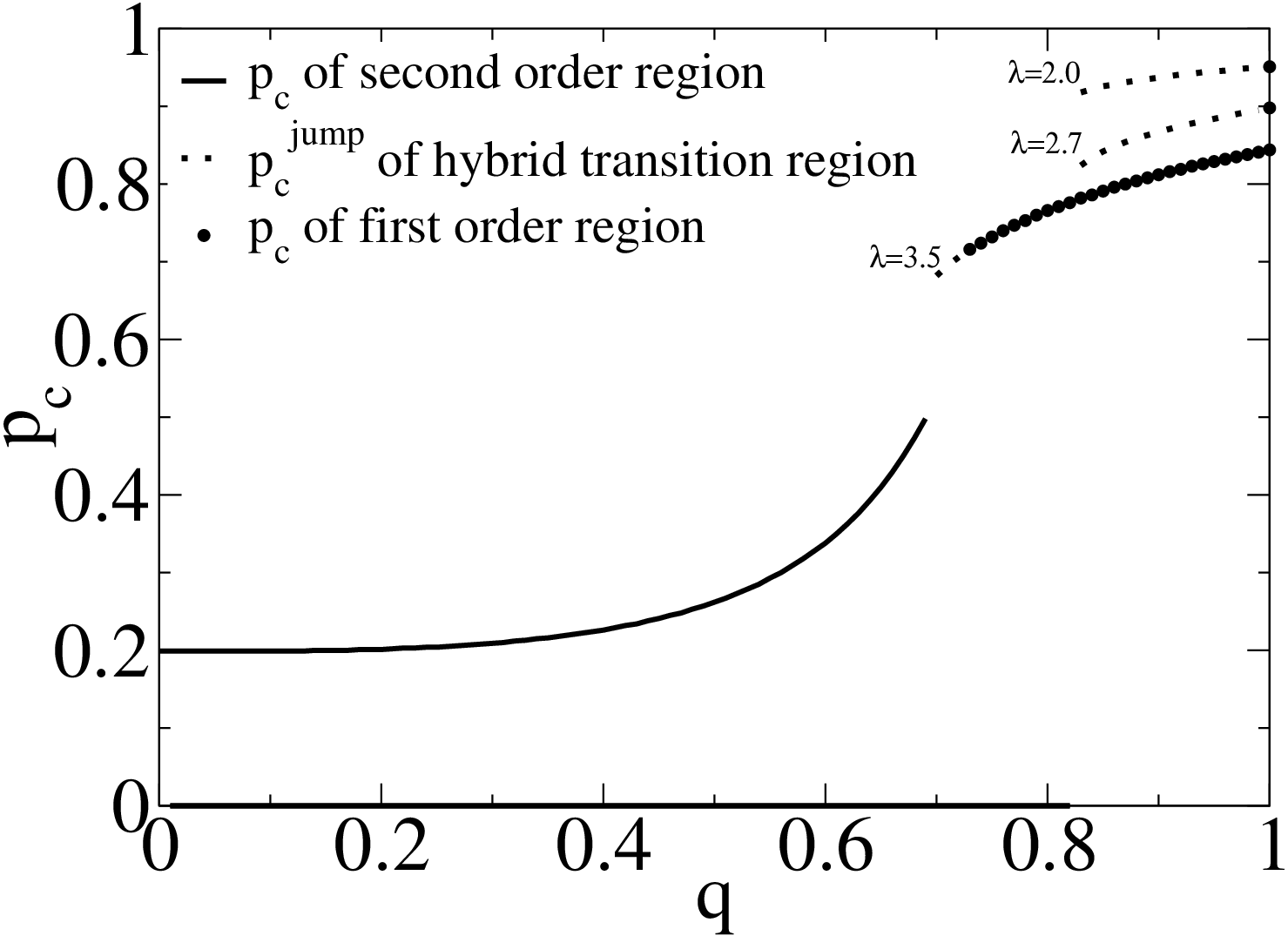}
\end{center}
\caption{(Color online) The critical threshold $p_c$ as a function of $q$ for $\lambda=2.0,~2.7$, and $3.5$.
The values of $p_c$ are
defined as follows: for the first-order transition,
$p_c$ is where the $\psi_{\infty}$ jumps to $0$; for hybrid transition,
$p_c^{jump}$ is where the sudden jump of $\psi_{\infty}$ to a non-zero $\psi_{\infty}$ occurs;
for second-order transition, $p_c$ is where $\psi_{\infty}$
goes to $0$. For $\lambda>3$, we can clearly see three regions of $p_c$.
For $\lambda<3$, $q_1=1$ and for $q<q_2\approx 0.83$, $p_c^{jump}$ disappears and
$p_c$ becomes zero.}
\label{Fig5}
\end{figure}

Now for any given $\lambda$ value, we plot in Fig.~\ref{Fig5}, $p_c$ as a
function of $q$ ($p_c(q)$). For $\lambda \leqslant 3$, only when $q=1$ it is a
first-order transition, where $\psi_{\infty}$ abruptly goes to $0$ below $p_c(1)$;
When $q<1$, it is either
hybrid or second-order transition, and $\psi_{\infty}$
is strictly $0$ only at $p=0$ for both cases.
However, since for the hybrid transition, the giant component becomes very small at
$p^{jump}_c$, we can regard this point as an effective $p_c$.
For second-order transition, although there still
exists a $p$ value where there is a maximum change in the magnitude of
$\psi_{\infty}$, but since
$\psi_{\infty}$ is continuous in all $p$ region, we define $p_c$ where
$\psi_{\infty}$ goes to $0$ and thus $p_c$ is always $0$.

For $\lambda>3$, the first-order transition
happens also for $q < 1$ and at $p_c$,
$\psi_{\infty}$ jumps to $0$.
In this case $p_c$ of the second-order
transition and of the hybrid transition is not $0$.

\section{Summary}
We find that for two SF interdependent
networks model with partial dependency $q$, the phase transition behavior of the giant
cluster under random attack shows a change from first-order (for $q_1 < q < 1$)
through hybrid transition ($q_2 < q < q_1$) to
a second-order phase transition ($0 \leqslant q < q_2$). In the hybrid
transition region, at an effective critical point $p^{jump}_c$,
the giant component $\psi_\infty$
has a sharp drop from finite value to a much smaller, yet a $non-zero$ value.
The hybrid transition seems to be unique for SF
since it does not appear in coupled ER networks \cite{Rony2010, ER1959, Bollobas1985}.

\section{Acknowledgement}
We thank DTRA and Office of Naval
Research for support. S. H. thanks the European LINC, EPIWORK and MULTIPLEX projects, 
the Deutsche Forschungsgemeinschaft
(DFG)and the Israel Science Foundation for financial support.

\bibliographystyle{apsrev}

\newpage

\end{document}